\PassOptionsToPackage{usenames}{xcolor}
\PassOptionsToPackage{dvipsnames}{xcolor}

\documentclass[sigconf]{acmart} 

\usepackage{cleveref}

\usepackage{subcaption}
\usepackage{tabularx}

\usepackage{enumitem}

\usepackage{booktabs} 

\usepackage{csquotes}
\usepackage{amsmath}
\usepackage{outlines}

\usepackage{eurosym}
\usepackage{siunitx}
\usepackage[usenames]{xcolor}

\usepackage{graphicx}

\usepackage{pifont}
\usepackage{array}
\newcolumntype{P}[1]{>{\centering\arraybackslash}m{#1}}
\newcolumntype{L}[1]{>{\arraybackslash}m{#1}}

\usepackage{xspace}

\newcommand\eg{e.\,g.\xspace}

\newcommand\US{U.\,S.\xspace}

\DeclareMathOperator{\logit}{logit}

\usepackage{url}
\usepackage{balance}

\makeatletter
\renewcommand{\fps@figure}{htb}         
\renewcommand{\fps@table}{htb}         
\makeatother

\allowdisplaybreaks

\AtBeginDocument{%
\providecommand\BibTeX{{%
\normalfont B\kern-0.5em{\scshape i\kern-0.25em b}\kern-0.8em\TeX}}}

\setcopyright{acmcopyright}
\copyrightyear{2022}
\acmYear{2022}
\acmDOI{}

\acmConference[WWW '22]{Lyon '22: ACM The Web Conf}{April 25--29, 2022}{Lyon, France}
\acmBooktitle{Proceedings of the ACM The Web Conf (WWW '22), April 25--29, 2022, Lyon, France}
\acmPrice{}
\acmISBN{}

\begin{document}

\fancyhead{}

\title{Hate Speech in the Political Discourse on Social Media: \\Disparities Across Parties, Gender, and Ethnicity}

\author{Kirill Solovev}
\email{kirill.solovev@wi.jlug.de}
\affiliation{
	\institution{JLU Giessen}
	\streetaddress{Licher Str.\ 62}
	\country{Germany}
}

\author{Nicolas Pröllochs}
\email{nicolas.proellochs@wi.jlug.de}
\affiliation{
	\institution{JLU Giessen}
	\streetaddress{Licher Str.\ 62}
	\country{Germany}
}
\renewcommand{\shortauthors}{Kirill Solovev and Nicolas Pröllochs}

\begin{abstract}
Social media has become an indispensable channel for political communication. However, the political discourse is increasingly characterized by hate speech, which affects not only the reputation of individual politicians but also the functioning of society at large. In this work, we empirically analyze how the amount of hate speech in replies to posts from politicians on Twitter depends on personal characteristics, such as their party affiliation, gender, and ethnicity. For this purpose, we employ Twitter's Historical API to collect every tweet posted by members of the 117th \US Congress for an observation period of more than six months. Additionally, we gather replies for each tweet and use machine learning to predict the amount of hate speech they embed. Subsequently, we implement hierarchical regression models to analyze whether politicians with certain characteristics receive more hate speech. We find that tweets are particularly likely to receive hate speech in replies if they are authored by (i) persons of color from the Democratic party, (ii) white Republicans, and (iii) women. Furthermore, our analysis reveals that more negative sentiment (in the source tweet) is associated with more hate speech (in replies). However, the association varies across parties: negative sentiment attracts more hate speech for Democrats (vs. Republicans). Altogether, our empirical findings imply significant differences in how politicians are treated on social media depending on their party affiliation, gender, and ethnicity.
\end{abstract}




\maketitle

\section{Introduction}


Social media has become an indispensable communication channel for politicians in the \US and around the world. Compared to traditional media, it provides a number of key benefits for politicians: (i) social media provides a tool to spread messages to the public at scale, thereby increasing people's awareness of their (political) agenda~\cite{Ross.2015,Graham.2013,Hong.2012}. (ii) Social media encourages the dialogue between politicians and users, allowing for direct feedback from constituents and discussions of political ideas~\cite{Enli.2013}. (iii) Due to its interactive nature, social media can be used as a tool for political mobilization~\cite{Jackson.2009,Larsson.2015}. These benefits are further reinforced by the openness of social media as politicians are no longer restricted by geography, scope, or content and can reach significantly wider audiences~\cite{Gainous.2013}. 

However, the shift from traditional channels towards social media does not necessarily improve the quality of the political discourse. Instead, social media is known to foster echo chambers and ``us versus them'' rhetoric~\cite{Mondal.2017}. These factors correlate with cyber-bullying, harassment, and, in particular, hate speech~\cite{Erjavec.2012}. Broadly speaking, hate speech refers to abusive or threatening speech (or writing) that expresses prejudice against a particular group, often on the basis of ethnicity or sexual orientation~\cite{Sellars.2016}. Hate speech often originates from semi-anonymous trolls~\cite{Wallace.2008,Mondal.2017}, and is particularly frequent in discussions that cause a strong emotional response, such as in political topics~\cite{Wagner.2020}. The adoption of social media by politicians is thus a double-edged sword posing risks both to themselves and society as a whole~\cite{Hong.2019}. At the individual level, hate speech can threaten reputations and may even lead to long-run mental health issues~\cite{Vidgen.2021}. At the societal level, it fosters political polarization~\cite{JamesA.Piazza.2020}, which can have severe consequences. Examples include erosion of intergroup political relations and increased opportunities for the spread of ideologically branded misinformation~\cite{Freelon.2020b,Prollochs.2022a,Solovev.2022b}.


\vspace{0.2cm}
\textbf{Research Goal:}
In this study, we empirically analyze how the user base on Twitter responds to posts from members of the \US Congress. We are interested in understanding whether differences in the prevalence of hate speech can be explained by personal characteristics of politicians, such as their party affiliation, gender, and ethnicity. More precisely, we address the following research questions:
\begin{itemize}[leftmargin=.5cm]
	\setlength\itemsep{.2em}
	\item \textbf{(RQ1)} \emph{Are Twitter users more likely to respond with hate speech to tweets from \US representatives depending on party affiliation, gender, and ethnicity of the members of the \US Congress?}
	\item \textbf{(RQ2)} \emph{Does hate speech in the replies to tweets depend on the sentiment of the source tweet? Does the strength of the association differ depending on their party, gender, and ethnicity?}
\end{itemize}

\newpage


\vspace{0.2cm}
\textbf{Data \& Methods: }
To address our research questions, we employ the Twitter Historical API to collect all tweets from members of the 117th \US Congress between the first session on January 3, 2021 and the end of July 2021. In addition, we collect replies to each source tweet. We then use machine learning to determine the share of replies of each tweet that embeds hate speech. Subsequently, we implement a multilevel binomial regression model with random effects to estimate whether Twitter users are more likely to respond with hate speech depending on the party affiliation, gender, and ethnicity of the politician that has posted the tweet. 

\vspace{0.2cm}
\textbf{Contributions: } To the best of our knowledge, this study is the first to empirically model how hate speech in replies to tweets from politicians depends on their personal characteristics (party affiliation, gender, ethnicity). All else being equal, we find that tweets are more likely to receive hate speech in replies if they are authored by (i) persons of color from the Democratic party, (ii) white Republicans, and (iii) women. As an additional contribution, our analysis reveals that more negative sentiment (in the source tweet) is associated with more hate speech (in replies). However, the association varies across parties: negative sentiment attracts more hate speech for Democrats (vs. Republicans). Altogether, our findings fuel new insights into ongoing discussions on political polarization on social media and highlight disparities in how politicians are treated depending on their  party affiliation, gender, and ethnicity

\section{Background}
\label{sec:related_work}

\textbf{Political communication on Twitter: }
The use of social media by \US politicians has experienced a rapid surge. At the start of 2009, only 69 individual members of Congress had a Twitter account~\cite{Golbeck.2010b}. Today, every member of the \US Congress has a professional Twitter account and oftentimes a second personal account being active at the same time. Existing studies suggest that there are three main reasons \emph{why} politicians adopt social media~\cite{Hong.2019}. First, social media allows for \emph{unidirectional delivery} of information to the public. Compared to classical media, there is less moderation and real time scrutiny allowing politicians to freely express themselves~\cite{Allcott.2017}. Second, social media enables \emph{dialogue} between politicians and the public. Politicians can use social media as a tool to connect with constituents to discuss political issues and receive feedback~\cite{Enli.2013}. Engaged users may further spread the message with likes and/or reshares. Third, social media can be seen as a tool for \emph{political mobilization}. Specifically, it allows politicians to rally for projects, events, and movements~\cite{Theocharis.2015}, though it does not guarantee success~\cite{Margetts.2015}.

\textbf{Hate speech: } Although there is no all-encompassing definition~\cite{Benesch.2014}, hate speech is typically considered to refer to abusive or threatening speech (or writing) that expresses prejudice against a particular group, often on the basis of ethnicity or sexual orientation~\cite{Sellars.2016}. While research on hate speech has received increasing attention lately~\cite[\eg,][]{Akhtar.2020, Chopra.2020, Davidson.2017, ElSherief.2018, Mossie.2020, Nagar.2021, Olteanu.2018, Saha.2021, Wich.2021, Zannettou.2018}, studies that analyze hate speech in the context of political communication are scant. The few existing works typically focus on qualitative insights or analysis of summary statistics. For instance, previous works have studied hate speech towards female Japanese politicians~\cite{Fuchs.2019}, far-right political party discourse in Spain~\cite{BenDavid.2016}, hateful propaganda towards politicians in Macedonia~\cite{BenDavid.2016}, hate speech against Members of Parliament in the U.K.~\cite{Agarwal.2021}, and hate against German politicians~\cite{Smedt.2018}. We are aware of only one paper analyzing hate speech and incivility in the context of tweets from members of the \US Congress~\cite{Theocharis.2020}. However, this study again focuses on summary statistics. In particular, it does not model the effects of personal characteristics of politicians (\eg, ethnicity) on the likelihood of receiving hate speech.

\textbf{Disparities across parties, gender, and ethnicity: }
Existing research suggest that political party leanings in the \US correlate with different speech patterns: Democrats tend to use more swear words and higher sentiment, while Republicans prefer to communicate more negative sentiment and group identity~\cite{Sylwester.2015}.
Besides party differences, a vast strand of studies has shown that there are discrepancies in communication behavior across genders. For instance, women are more likely to hide expressive and negative emotions~\cite{Davis.1995}, and are guided by a greater focus on care in moral dilemmas~\cite{Nguyen.2008}. This is directly applicable to the domain of social media, where women are more likely to report messages targeting racial minorities and women~\cite{Downs.2012}.
Gender differences are further reinforced by widespread stereotypes regarding the role of women in society~\cite{Prentice.2003}, who are perceived as less persuasive and are often outright dismissed when displaying aggressive and forceful behavior online~\cite{Winkler.2017}.
Furthermore, survey studies suggest that women more often tend to be a target of cyber-bullying and hateful attacks~\cite{Beckman.2013}, especially if they present an openly active stance, such as feminism~\cite{Hardaker.2016}.
Ethnicities and racial stereotypes play a similar role in offline and online discourse and differ greatly across countries~\cite{Tatum.2001}.
For instance, for the \US, existing studies suggest frequent hate speech against African Americans~\cite{Kwok.2013}.

\textbf{Research gap: } Existing research on hate speech in the political discourse focuses either on qualitative insights or on summary statistics. We are not aware of previous works empirically modeling the effect of personal characteristics on the likelihood of a politician to receive hate speech. This presents our contribution.

\section{Dataset}

\begin{figure}
	\centering
	\includegraphics[width=0.82\columnwidth]{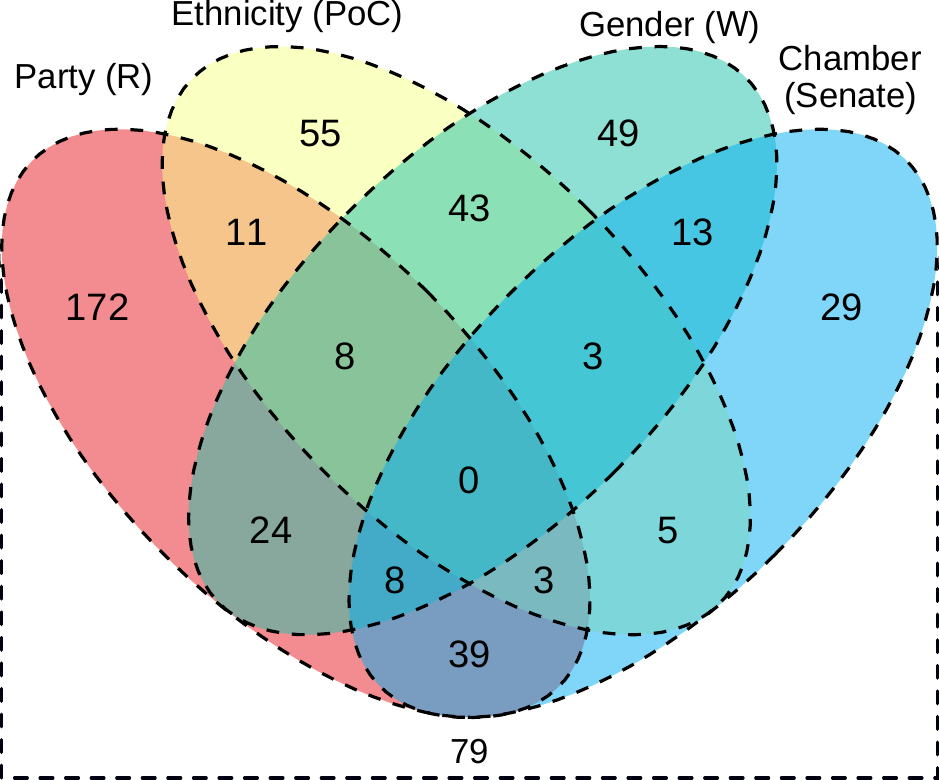}
	\caption{Venn Diagram visualizes the composition of the 117th \US Congress.}
	\label{fig:venn_congress}
\end{figure}

\textbf{Members of the \US Congress: } 
We analyze tweets from all 541 members of the 117th \US Congress that convened on January 3, 2021. 
Data on the members of Congress was gathered from the official webpage of the \US Congress~\cite{Congress.2021}, which provides links to personal and campaign web pages. By following these links, we collected the following information about each politician: (i) party affiliation, (ii) branch of Congress in which the politician serves, (iii) time served in Congress, (iv) gender, and (v) ethnicity.
Fig.~\ref{fig:venn_congress} provides an overview of the composition of the 117th \US Congress. 
Most voting seats are held by members of the two major political parties with 269 Democrats~(D) and 263 Republicans~(R), while 2 seats are occupied by independent senators.
Women~(W) hold 27\% of all Congress Seats, accounting for 39\% of all Democrats and 15\% of all Republicans, respectively.
Notably, the 117th \US Congress is the most ethnically diverse so far with 39\% of Democrats and 8\% of Republicans identifying as people of color~(PoC).

For the sake of simplicity and interpretability, we focus our later empirical analysis on tweets from Republican and Democratic members; and exclude tweets from the two independent senators.

\textbf{Collection of tweets: }
Twitter handles (user names) of every politician in the \US Congress are provided by the University of California San Diego library~\cite{Smith.2021}. We employed the Twitter Historical API to download the complete timelines of every politician between January 3, 2021 and the end of July 2021.
Here we collected the entire tweet history of each person, excluding retweets and replies, resulting in a total number of 199,294 tweets. The average number of tweets per politician is 368.38.
We additionally queried Twitter's Historical API to gather the replies to every source tweet in our data set.
To ensure feasibility, we restricted the data collection to up to 250 replies for each original tweet, starting with the earliest reply. The crawling process resulted in a total number of 8,362,555 replies. 

\section{Methods}
\label{sec:methods}

\subsection{Hate Speech Detection}

In this work, we use machine learning to detect hate speech in replies to tweets. Compared to dictionary-based methods that merely count hate-related words~\cite{Alrehili.2019}, this approach is generally considered as being more accurate~\cite{Badjatiya.2017}. Nonetheless, as part of our robustness checks, we validate our results with the frequently-employed Hatebase dictionary~\cite{Hatebase.2021}, finding confirmatory results.

We implement machine learning for hate speech detection as follows: we employ the annotated twitter dataset from~\cite{Davidson.2017}, containing 25,000 tweets labeled as hateful or not hateful. Each tweet was annotated by at least 3 users who were explicitly instructed to think about the context of the message and not only the words contained within~\cite{Davidson.2017}.
We use the annotated tweets to implement a deep neural network classifier that predicts whether or not a tweet is hateful.\footnote{{We use Universal Sentence Encoder (USE)~\cite{Cer.2018} as text representation. The machine learning classifier yields a weighted out-of-sample $F1$ score of 0.89, which is similar to previous works~\cite{Davidson.2017} and can be seen as reasonably accurate in the context of our study. The model is implemented in Python 3.8.5 using TensorFlow 2.6.0~\cite{MartnAbadi.2015}.}}
The hate speech classifier is then used to predict a binary label of whether or not a tweet is hateful ($=1$ if true; otherwise $=0$) for each reply tweet in our dataset.
For each source tweet, we calculate the share of replies that are hateful. The resulting variable ranges from 0 to 1, with 0 indicating the lack of hate speech in replies, and 1 indicating that every reply is hateful.

\subsection{Explanatory Regression Model}

We implement a multilevel binomial regression to estimate the effects of party, gender, and ethnicity on the likelihood of a tweet receiving hate speech. 

Formally, we model the number of hate speech replies, $HReplies$, as a binomial variable with probability parameter $\theta$. The number of trials is given by the total number of replies a tweet receives ($Replies$). The key explanatory variables are the politicians' party affiliation ($\mathit{Party}$; $=1$ if Republican, otherwise 0), gender ($\mathit{Gender}$; $=1$ if Man, otherwise 0), and ethnicity ($\mathit{Ethnicity}$; $=1$ if Person of Color, otherwise 0). Furthermore, for each source tweet, we calculate a sentiment score ($\mathit{SourceSentiment}$) using \texttt{SentiStrength}. 
We also control for the congressperson's age ($\mathit{Age}$), the number of years served ($\mathit{YearsInOffice}$), whether media was attached to the tweet ($\mathit{AttachedMedia}$; $=1$ if true, otherwise 0), and the chamber of congress at which the politician serves ($\mathit{Chamber}$; $=1$ if Senate, otherwise 0). Based on these variables, we specify the following regression model:
\begin{flalign}
	\logit(\theta) = & \, \beta_0 + \beta_{1} \mathit{Party} + \beta_{2} \mathit{Gender} + \beta_{3} \mathit{Ethnicity} \label{eq:theta} \\
											& + \beta_{4} \mathit{SourceSentiment} + \beta_{5} \mathit{YearsInOffice} +  \beta_{6} \mathit{Age} \nonumber \\
                                            & + \beta_{7} \mathit{AttachedMedia} + \beta_{8} \mathit{Chamber} \nonumber \\
											& + u_\text{user} + \varepsilon \text{,} \nonumber
\end{flalign} \vspace{-1\baselineskip}
\begin{flalign}
    \hspace{1ex} HReplies \sim Binomial[Replies, \theta], & \hskip29mm \label{eq:binom}
\end{flalign}
with intercept $\beta_0$, error term $\varepsilon$, and user-specific random effects $u_\text{user}$. Note that the latter is important as it allows us to control for heterogeneity in users' social influence (\eg, some accounts have many followers and reach different audiences) \cite{Prollochs.2021b,Prollochs.2021a}. 

We estimate Eq.~\ref{eq:theta} and Eq.~\ref{eq:binom} using MLE and generalized linear models. To facilitate the interpretability of our findings, we $z$-standardize all variables, so that we can compare the effects of regression coefficients on the dependent variable measured in standard deviations.
Our regression analyses are implemented in R~4.0.5 using the \texttt{lme4} package~\cite{Bates.2007}. 

\section{Empirical Analysis}

\subsection{Summary Statistics}
\begin{figure}
	\captionsetup{position=top}
	\centering
	\subfloat[]{\includegraphics[width=.42\linewidth]{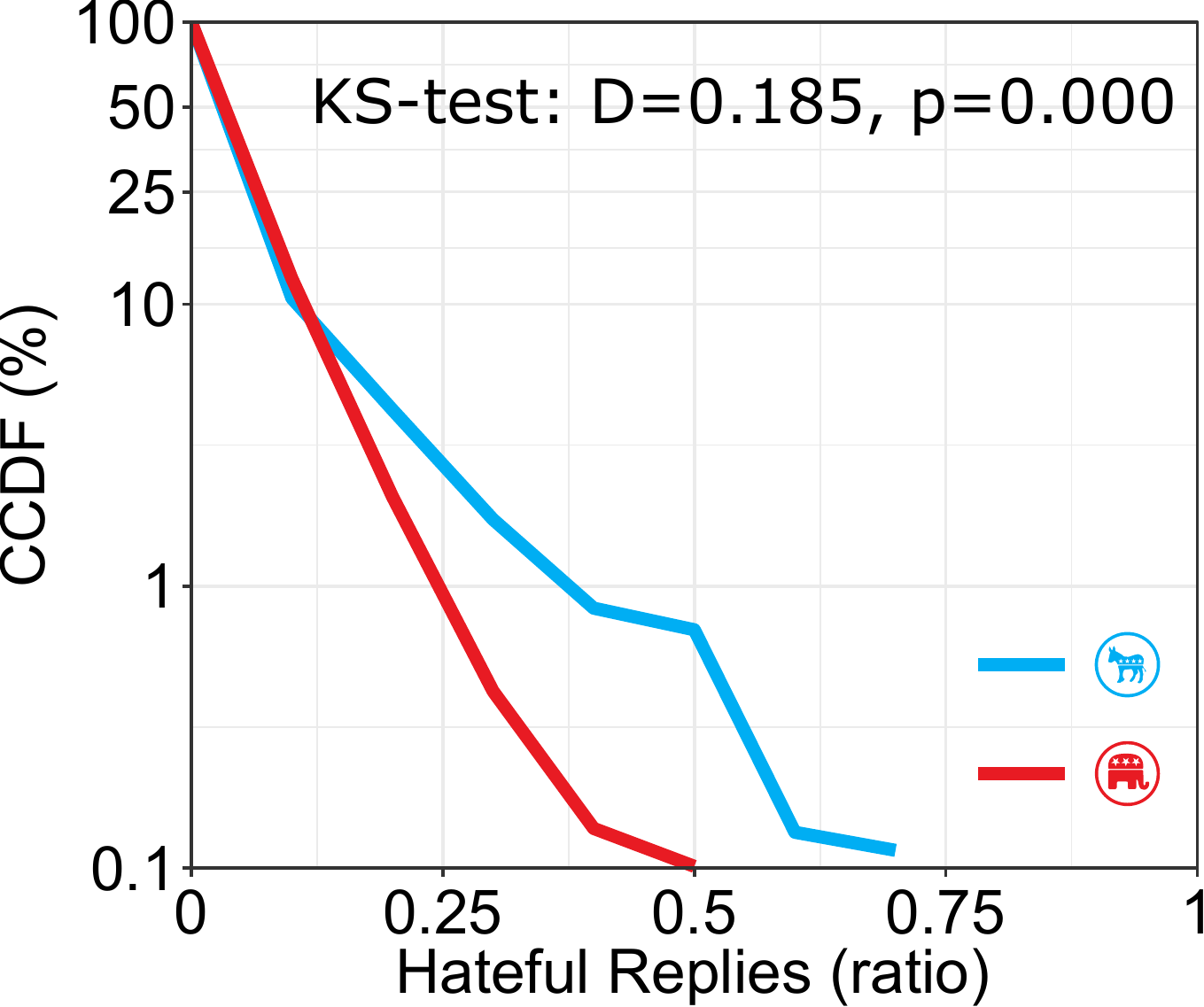}\label{fig:ccdf_hate_party}}
	\hspace{0.4cm}
	\subfloat[]{\includegraphics[width=.42\linewidth]{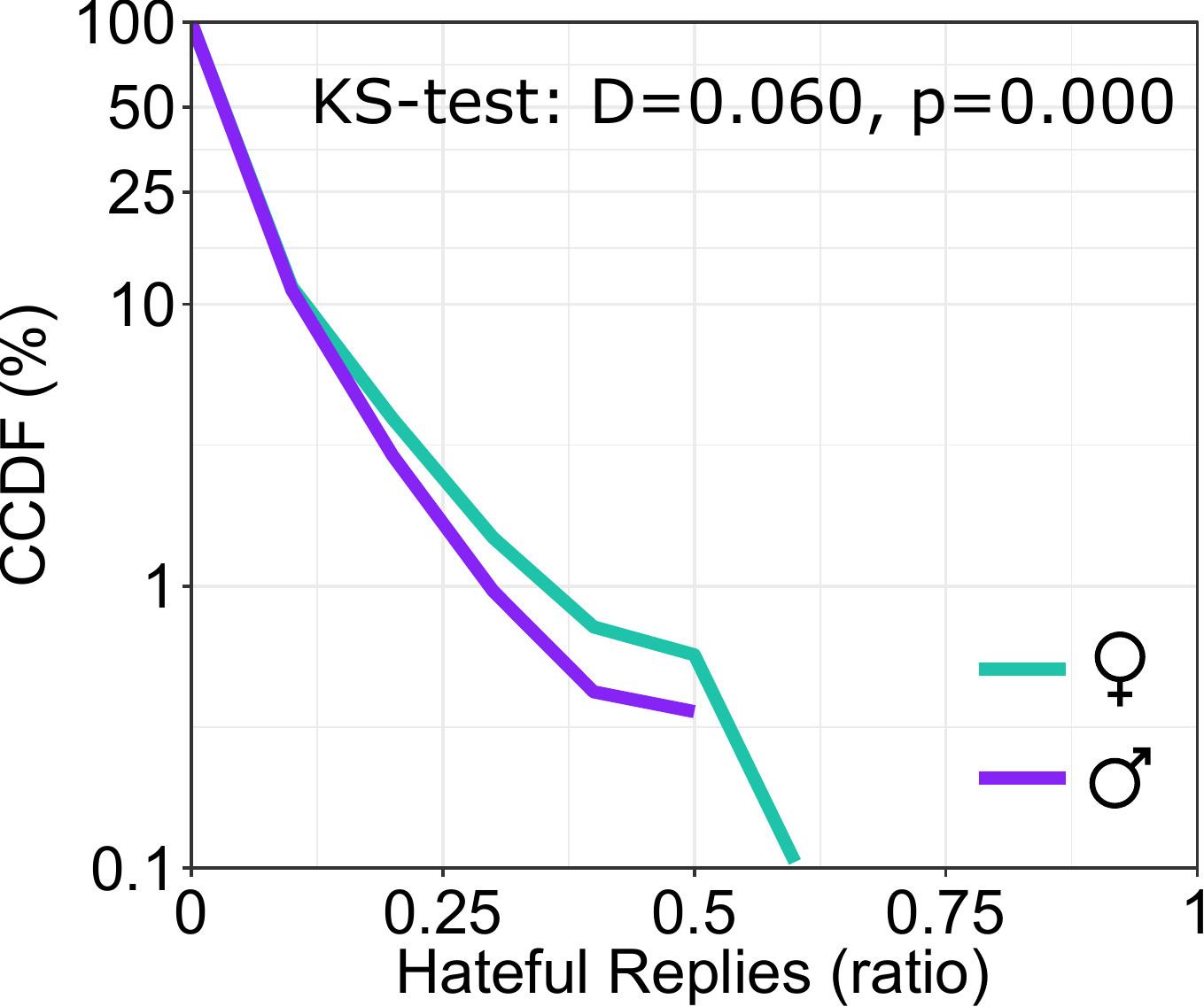}\label{fig:ccdf_hate_gender}}
	\hspace{0.4cm}
	\subfloat[]{\includegraphics[width=.42\linewidth]{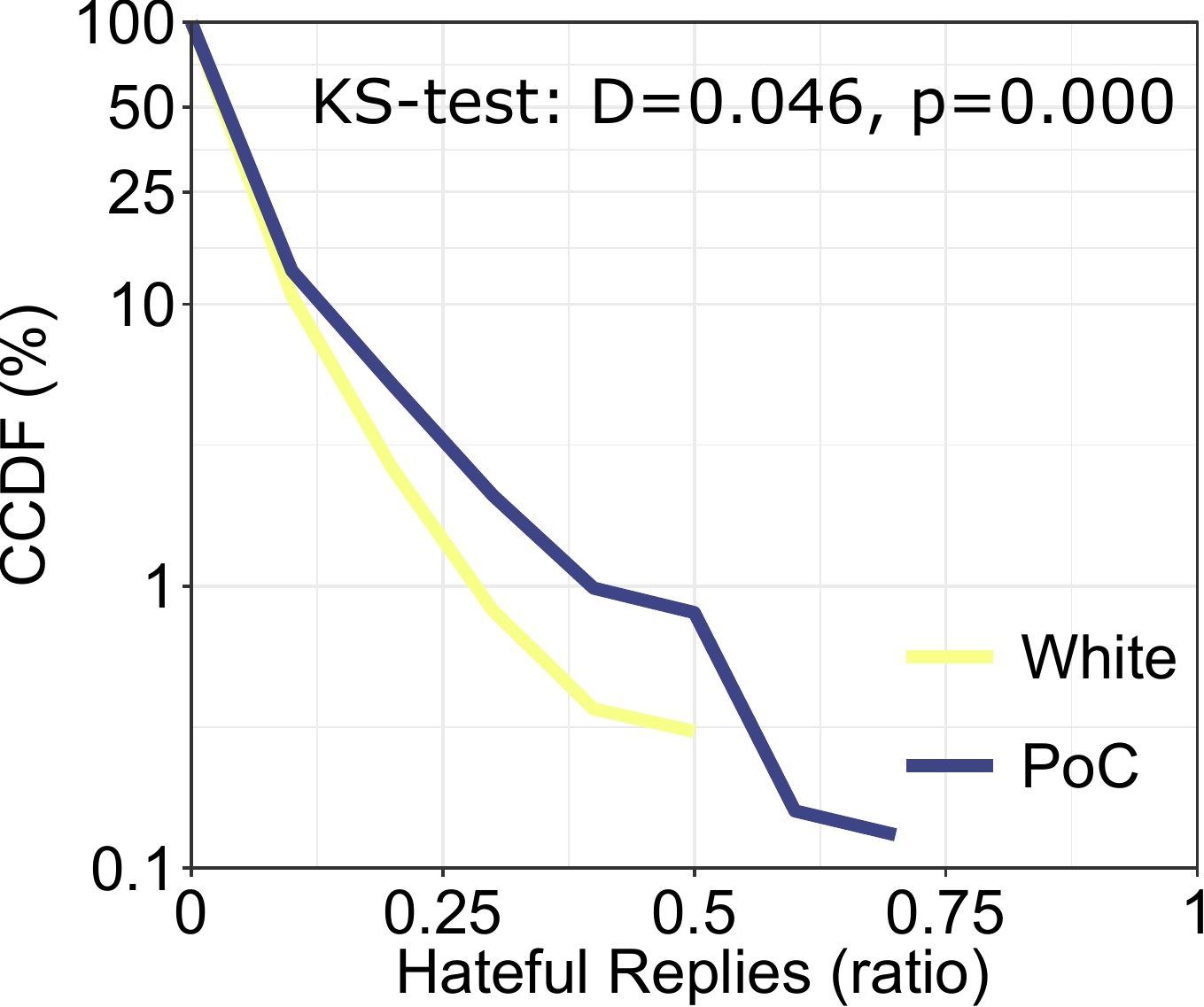}\label{fig:ccdf_hate_poc}}
	\vspace{.25cm}
	\caption{CCDFs for the ratio of hate speech in replies separated by (a) party, (b) gender, and (c) ethnicity.}
	\label{fig:ccdfs_replies}
\end{figure}

We start our analysis by evaluating summary statistics. 
The average share of hateful replies per tweet in our dataset amounts to \SI{1.99}{\percent}. We perform both $t-$tests and Kolmogorov-Smirnov (KS) tests to evaluate whether there are statistically significant differences across parties, genders, and ethnicities. Our findings are as follows: (i)~tweets from Democrats (vs. Republicans) receive, on average, a 3.67\% higher share of hate replies. (ii)~Tweets from women (vs. men) politicians receive 7.71\% higher share of hate replies. (iii)~Tweets from persons of color (vs. whites) receive 37.75\% higher share of hate replies. For each of these comparisons, two-sided $t-$tests confirm that the differences in means are statistically significant ($p<0.01)$. In Fig.~\ref{fig:ccdfs_replies}, we visualize the complementary cumulative distribution functions (CCDFs) for the ratio of hate speech in replies. We again find that Democrats, women and persons of color receive more hate speech. KS-tests confirm that all differences in distributions are statistically significant ($p<0.01$).

\subsection{Regression Analysis}

\begin{figure}
	\centering
	\includegraphics[width=\columnwidth]{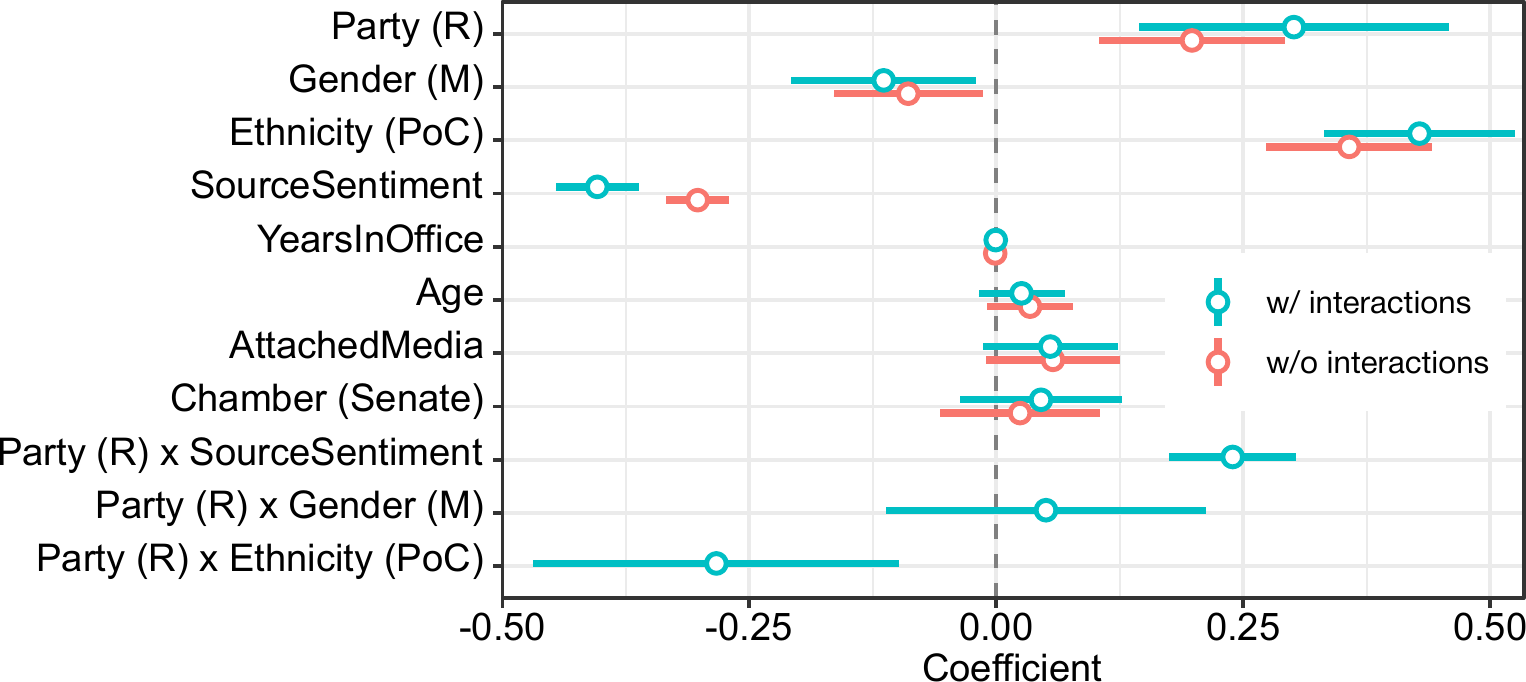}
	\caption{Coefficient estimates for binomial regression w/o (coral) and w/ (teal) interaction terms for political party. The horizontal bars represent 95\% confidence intervals. User-specific random effects are included.}
	\label{fig:coef_hate_bin}
\end{figure}

We estimate a multilevel binomial regression to understand the effects of party affiliation, gender, and ethnicity on the likelihood of a tweet receiving hate speech (see model w/o interactions in Fig.~\ref{fig:coef_hate_bin}). In contrast to summary statistics, this allows us to estimate effect sizes \emph{after} controlling for confounding effects. The largest effect size is estimated for \textit{Ethnicity} with a coefficient of \num{0.346} ($p<0.01$), which implies that the odds of receiving hate speech for persons of color are $e^{0.346} \approx 1.41$ times the odds for whites. We further observe pronounced party and gender effects. Compared to Democrats, the odds for tweets from Republicans to receive hate speech are 22.02\% higher ($\beta = 0.199$, $p < 0.01$). The odds for men to receive hate speech are 8.33\% ($\beta = -0.087$, $p < 0.05$) lower than for women. We also find that a more negative sentiment in the source tweet is associated with more hate speech in replies. A one standard deviation increase in \textit{SourceSentiment} is associated with a 25.99\% ($\beta = -0.301$, $p < 0.01$) decrease in the odds of receiving hate speech. We find no statistically significant effects from a politician's age, time in office, chambers, and media attachments.


We add interaction terms to test whether users react differently to gender, ethnicity, and sentiment depending on the party affiliation (see model w/ interactions in Fig.~\ref{fig:coef_hate_bin}). Here we find a statistically significant interaction term between $\textit{Party}$ and $\mathit{Ethnicity}$ ($\beta = -0.287$, $p < 0.01$). This implies that persons of color from the Democratic party have higher odds for receiving hate speech than persons of color from the Republican party. Furthermore, the strength of the association between sentiment in the source tweet and hate speech varies across parties ($\beta = 0.235$, $p < 0.01$). Specifically, negative sentiment attracts more hate speech for Democrats. The interaction between party affiliation and gender is not significant at common statistical significance thresholds.

Altogether, our analysis implies that three groups of politicians are particularly likely to receive hate speech in response to their tweets: (i) persons of color from the Democratic party, (ii) white Republicans, and (iii) women.

\subsection{Robustness Checks}

We conducted additional checks to validate the robustness of our analysis: (1) We repeated our analysis with a dictionary-based approach for hate speech detection, specifically the Hatebase dictionary~\cite{Hatebase.2021}. (2)~We calculated variance inflation factors for all independent variables in our regression model and found that all remain below the critical threshold of four.
(3)~We repeated our analysis with alternative estimators (\eg, beta regression), controlled for outliers, tested for quadratic effects, and added multiple interaction terms for each explanatory variable.
In all cases, our results are robust and consistently support our findings.

\section{Discussion}


\textbf{Summary of findings: }
This work empirically models how the amount of hate speech in replies to tweets from politicians depends on their personal characteristics (party affiliation, gender, ethnicity). All else being equal, we find that Tweets are particularly likely to receive hate speech replies if they are authored by (i) persons of color from the Democratic party, (ii) white Republicans, and (iii) women. Furthermore, our analysis reveals that more negative sentiment (in the source tweet) is associated with more hate speech (in replies). However, the association varies across parties: negative sentiment attracts more hate speech for Democrats (vs. Republicans). Altogether, our empirical findings imply statistically significant differences in how politicians are treated on social media depending on their party affiliation, gender, and ethnicity.


\textbf{Implications: }
Our findings are relevant both for politicians and from a societal perspective. Politicians should be aware that social media is a double-edged sword as it comes with the risk of receiving vast numbers of hate comments. This is concerning as hate speech can destroy reputations and may even lead to long-run mental health consequences~\cite{Vidgen.2021}. Given that hate speech can affect peoples' decision to participate in politics~\cite{Scott.2019}, this may also impede diversity in the composition of political institutions. Furthermore, hate speech goes hand in hand with increased polarization, hyper-partisanship, and less common ground between opposing political sides ~\cite{Finkel.2020}, thereby threatening the functioning of democracy itself.

\bibliographystyle{ACM-Reference-Format-no-doi-abbrv}
\balance
\bibliography{literature}

%
%
%
%

\end{document}